**Probing Local Variations of Superconductivity on the Surface of Ba(Fe$_{1-x}$Co$_x$)$_2$As$_2$ Single Crystals**

T.-H. Kim[1], R. Jin[2], L. R. Walker[3], J. Y. Howe[3], M. H. Pan[1], J. F. Wendelken[1], J. R. Thompson[3,4], A. S. Sefat[3], M. A. McGuire[3], B. C. Sales[3], D. Mandrus[3], and A. P. Li[1,4*]

[1] Center for Nanophase Materials Sciences, Oak Ridge National Laboratory, Oak Ridge, TN 37831, USA

[2] Department of Physics & Astronomy, Louisiana State University, Baton Rouge, LA 70803, USA

[3] Materials Science and Technology Division, Oak Ridge National Laboratory, Oak Ridge, TN 37831, USA

[4] Department of Physics and Astronomy, University of Tennessee, Knoxville, Tennessee 37996, USA

The spatially resolved electrical transport properties have been studied on the surface of optimally-doped superconducting Ba(Fe$_{1-x}$Co$_x$)$_2$As$_2$ single crystal by using a four-probe scanning tunneling microscopy. While some non-uniform contrast appears near the edge of the cleaved crystal, the scanning electron microscopy (SEM) reveals mostly uniform contrast. For the regions that showed uniform SEM contrast, a sharp superconducting transition at $T_C$ = 22.1 K has been observed with a transition width $\Delta T_C$ = 0.2 K. In the non-uniform contrast region, $T_C$ is found to vary between 19.6 and 22.2 K with $\Delta T_C$ from 0.3 to 3.2 K. The wavelength dispersive x-ray spectroscopy reveals that Co concentration remains 7.72% in the uniform region, but changes between 7.38% and 7.62% in the non-uniform region. Thus the variations of superconductivity are associated with local compositional change.

---

* Corresponding author. apli@ornl.gov

*PACS* numbers: 74.25.Fy, 74.62.Bf, 07.79.Fc

# I. INTRODUCTION

The discovery of an iron-arsenide family of high-$T_C$ superconductors [1-4] has aroused intensive experimental effort to delineate the nature of superconductivity. In particular, it has often been compared with the high-$T_C$ superconductivity in cuprates [5]. These materials indeed share many common features. Both materials are layered with electronically active Cu-O and Fe-As planes alternating with buffer layers of different chemical compositions. However, there exist clear differences between these two types of superconducting materials too. First, the Fe-As layers in Fe-based superconductors are formed by edge-shared $FeAs_4$ tetrahedra, instead of the corner-shared Cu-O layers with Cu in the octahedral environment in high-$T_C$ cuprates. Second, the ground state of undoped parent compounds of Fe-based superconductors is metallic without Mott-insulator transition, even though it has antiferromagnetic order and undergoes a structural phase transition. Of most prominence is the great chemical flexibility of this new family with almost all elements replaceable. For example, superconductivity in $BaFe_2As_2$ can be induced by doping in either Ba sites [4] or Fe sites, with either magnetic or non-magnetic elements [6, 7]. The latter is in distinct contrast to the cuprates, where even dilute substitution with non-magnetic elements into the Cu-O planes will suppress $T_C$ and eventually destroy the superconductivity. These differences suggest that the underlying physics in these two families of superconductors may be different.

Recently, the combined real-space imaging of scanning tunneling microscopy/spectroscopy (STM/STS) and momentum space diffraction have revealed that the cleaved surface of $BaFe_2As_2$ single crystals is an ordered Fe-As plane [8], although Ba induced stripe features have also been observed in doped phases [9, 10]. This thus provides an ideal surface for studying the interplay between superconductivity and dopants in the Fe sites using surface sensitive techniques. In this article, we present a spatially resolved superconductivity study on superconducting Ba(Fe$_{1-x}$Co$_x$)$_2$As$_2$ single crystals for the first time, using a combination of four-probe STM and scanning electron microscopy (SEM). We find that SEM images of cleaved surfaces of Ba(Fe$_{1-x}$Co$_x$)$_2$As$_2$ single crystals show mostly uniform contrast, while some non-uniform contrast appears near the edge of the crystal. In the uniform regions, the four-probe STM measurements reveal a sharp superconducting transition, showing superconductivity below $T_C$ = 22.1 K and transition width $\Delta T_C$ = 0.2 K. In the non-uniform regions, variations of both $T_C$ and $\Delta T_C$ are observed. Quantitative analysis of Co distribution revealed by wavelength dispersive x-ray spectroscopy (WDS) indicates that the variations of both $T_C$ and $\Delta T_C$ are related to the change of Co concentration in the non-uniform region. This suggests that the superconductivity is coupled with the local environment, although charge carriers are itinerant in nature.

## II. EXPERIMENTAL DETAILS

Single crystals of Co-doped BaFe$_2$As$_2$ were grown out of FeAs flux, and characterized by x-ray diffraction and electron-probe micro-analysis (EPMA). Elemental analysis of the crystals was performed using WDS (JEOL JXA-8200 Superprobe). The bulk crystals show superconductivity below 22 K with a transition width of 0.6 K. Further details about the sample preparation and the bulk superconducting properties have been reported in ref. [7]. For the local electrical transport study, crystals were cleaved in ultra-high vacuum (UHV) of < 2×10$^{-10}$ Torr at room temperature and immediately transferred into the UHV four-probe STM chamber, where they can be cooled to 10 K. The cleaved surfaces were examined by field-emission SEM and cryogenic STM. The four-point contact transport measurements were carried out in situ using four electrochemically etched STM (tungsten) probes that can be brought onto the same sample area within sub-micrometer scale guided by SEM [11]. Using a controlled tunneling feedback mechanism, the probes may then be precisely placed on a surface to permit transport measurements [11, 12]. The SEM beam was blanked off during transport measurements. A calibrated silicon diode (Lakeshore, DT-670-SD-4D) was mounted directly on the sample holder adjacent to the single crystal in order to measure sample temperature precisely. Typical accuracy of measured temperature at 20 K is ± 14 mK.

## III. RESULTS AND DISCUSSION

Single crystals of Ba(Fe$_{1-x}$Co$_x$)$_2$As$_2$ materials form layered structures and can be easily cleaved to expose large flat surfaces. The topographic images of cleaved surface of optimally doped BaFe$_{1.8}$Co$_{0.2}$As$_2$ (nominal composition will be presented for convenience) are shown in Fig. 1. Very flat terraces with *uniform* contrast have been observed in the secondary electron emission images acquired by SEM [Fig. 1(a)]. However, small darker regions in the SEM image have also been identified as marked by a rectangular box in Fig. 1(a). An SEM zoom-in image in the dark region reveals microscopic *domain* structures [Fig. 1(b)]. The SEM image in the secondary electron emission mode can register the contrast according to topography, chemical composition, and surface barrier (work function or ionization energy) of the sample [13]. As shown below, the contrast here reflects the local changes of chemical composition in the crystal. Figure 1(c) shows a topographic STM image taken in a uniform region at room temperature in constant current mode. It reveals an atomically flat surface with single atomic step height of 1.3 nm, confirming the *ab* cleavage plane. The quasi-atomic resolution image shown in the inset of Fig. 1(c) resembles previously reported one-dimensional stripe-like structural order in optimally-doped $BaFe_{1.8}Co_{0.2}As_2$ [9, 10].

Local transport *I-V* curves have been measured at different locations at variable temperatures by using a collinear four-probe method [12]. Figure 2 shows temperature-dependent resistance measured both in uniform contrast regions [measurement locations marked in Fig. 1(a)] and domain regions [marked in Fig. 1(b)]. In the uniform regions, the superconducting transition occurs at $T_C = 22.1$ K for 10% fixed percentage of the normal state resistance. The transition width is $\Delta T_C = T_C(90\%) - T_C(10\%) = 0.2$ K. In the domain regions, although the onset superconducting transition temperature (90% of the normal resistance) is very close to that of the uniform regions, $\Delta T_C$ varies over a broader range of 0.3 to 3.2 K. Furthermore, resistance of the domain regions above the transition onset temperature is higher than that of the uniform region, indicating higher defect density in the domain regions. Especially, in some locations, a residual resistance persists far below the onset of the superconducting transition temperature, suggesting the coexistence of superconducting and non-superconducting phases.

The critical current $I_C$, namely the maximum current allowed while maintaining the zero-resistance state, varies from location to location, too. Figure 3(a) shows the temperature dependence of $I_C$ in the critical state at four different sample locations: I and II from uniform regions, III and IV from the domain regions. For I and II, the transition temperatures are almost the same, though $I_C$ of II is slightly higher than that of I at the same temperatures. $I_C$ is generally higher in uniform regions than that of the domain regions, suggesting a stronger pinning effect in the uniform region and the inhomogeneity of superconductivity across the surface.

Furthermore, the temperature dependence of critical current shows an intriguing scaling behavior. As shown in Fig. 3(a), all $I_C(T)$ data both from the uniform and domain regions near $T_C$ on the surface follow a generalized power law, $I_C(T) = I_C(0)(1-T/T_C)^n$, giving the best fit with a critical exponent $n = 0.60 \pm 0.03$. The extrapolated $I_C(0) = 19 \pm 3$ mA in the uniformed regions, which is much higher than $I_C(0) = 8 \pm 1$ mA in the domain regions. Note that all data can be well superimposed onto a single curve of $I_C(T)/I_C(0) = (1-T/T_C)^n$ when $I_C(T)$ is normalized by $I_C(0)$ [shown in the inset of Fig. 3(a)]. A similar power law scaling over a wide temperature range has also been observed in the critical current density $j_C(T)$ derived from bulk remanent magnetization using the Bean model [shown in Fig. 3(b)], but with a higher critical exponent of $n = 1.46$. The scaling law of bulk crystal ($n = 1.46$) fits well to the Ginzburg-Landau (GL) critical current behavior $j_C \propto [1-T/T_C]^{3/2}$ [14-16], a good agreement with the weak-coupling BCS theory for a homogeneous order parameter. The reduced critical exponent from bulk 3/2 value on the surface may be attributed to following two factors. First, it is known that in superconductors with a thickness approaching the London penetration depth $\lambda$ (~200 nm at $T = 0$ K for $Ba(Fe_{0.93}Co_{0.07})_2As_2$ [17]), the finite thickness effect or edge effect will lead to a reduced temperature exponent [16, 18, 19]. If considering the surface nature of four-probe STM characterization [20] and the fact that the London penetration depth can be considerably greater near $T_C$ [17], the reduced exponent observed by four-probe near $T_C$ may be attributed to an edge effect. Secondly, non-GL scaling behavior has been reported in many high-$T_C$ cuprates, owing to weak links in the granular materials [15]. Even when superconductors are not granular in a structural sense, local variations of dopants can result in coexistence of superconducting and non-superconducting domains, and the crystal would exhibit granular superconductivity [21]. In the single crystals of $BaFe_{1.8}Co_{0.2}As_2$, the local depression of the superconducting state could be weak links as it moves away from optimal doping towards the competing parent antiferromagnetic phase.

The local composition analyses by WDS both with line scanning and two-dimensional elemental mapping have indeed shown a variation of chemical compositions across the surface. Figure 4 shows the spatially resolved Co concentrations both in the uniform and domain regions where local transport measurements were carried out. In the uniform contrast regions, Co distribution is uniform with approximately 7.72 % of Fe substituted by Co in the crystal. However, in the domain regions, the average Co content along a line [marked with dash line in Fig. 1(b)] increases gradually from 7.38% to 7.62%. Both quantitative line scans and semi-quantitative data extracted from elemental maps indicate the Co concentration increases along the line, suggesting the inhomogeneity of Co distributions in the domain regions. The local resistance has been probed using the four-probe STM following the same line at 21.1 K, which is near the superconducting transition temperature. As shown in Fig. 5, the resistance gradually decreases with increasing Co content. At 21.1 K, the right side of the region, having a Co content $\geq 7.61\%$, starts to show superconducting behavior but the left side with slightly lower Co content still shows a finite residual resistance. In contrast, the resistance in the uniform region remains negligibly small, indicating the persistence of the superconducting state. Given the fact that the normalized $I_C$ show a power law scaling with the same critical exponent in both uniform and non-uniform regions, the reduced surface critical current exponent compared to the bulk can thus be attributed to the edge effect rather than the granularity. On the other hand, it has been reported that superconductivity emerges for Co content above 2.5%, reaching a maximum $T_C$ around $x$ ~ 8% and decreasing $T_C$ for higher $x$ values [22, 23], qualitatively consistent with our observations. The quantitative correlation between the superconductivity and Co concentration confirms that the variations of superconductivity are a consequence of local change of the chemical composition in the crystal.

## IV. CONCLUSIONS

In summary, the local electrical transport measurements show that the superconducting behavior of an optimally doped $BaFe_{1.8}Co_{0.2}As_2$ is not uniform across the single crystal surface. The microscopic variations of superconductivity strongly correlate with local Co doping concentration. In fact, both cuprates and pnictides have a generic phase diagram, containing the ubiquitous superconducting "dome" around optimal doping. As a result, local composition change and the associated superconductivity variations have been seen in cuprates [21, 24]. In pnictide $BaFe_{1.8}Co_{0.2}As_2$, superconducting gap magnitude variations of about ±20% have been observed by STM/STS studies [9, 10], and a paramagnetic broadening of the NMR line has recently been observed, too [22]. Both these observations were believed to be effects of the local variations of Co dopants. Our spatially resolved electrical transport measurements have now provided direct evidence of the coupling between superconductivity and local environment that is reflected by Co concentration variation.

**ACKNOWLEDGMENTS**

[1] X. H. Chen, T. Wu, G. Wu, R. H. Liu, H. Chen, and D. F. Fang, Nature **453**, 761 (2008).

[2] Y. Kamihara, T. Watanabe, M. Hirano, and H. Hosono, J. Am. Chem. Soc. **130**, 3296 (2008).

[3] H. Kito, H. Eisaki, and A. Iyo, J. Phys. Soc. Jpn. **77**, 063707 (2008).

[4] M. Rotter, M. Tegel, and D. Johrendt, Phys. Rev. Lett. **101**, 107006 (2008).

[5] M. Norman, Physics **1**, 21 (2008).

[6] A. Leithe-Jasper, W. Schnelle, C. Geibel, and H. Rosner, Phys. Rev. Lett. **101**, 207004 (2008).

[7] A. S. Sefat, R. Jin, M. A. McGuire, B. C. Sales, D. J. Singh, and D. Mandrus, Phys. Rev. Lett. **101**, 117004 (2008).

[8] V. B. Nascimento, A. Li, D. R. Jayasundara, Y. Xuan, J. O'Neal, S. Pan, T. Y. Chien, B. Hu, X. B. He, G. Li, A. S. Sefat, M. A. McGuire, B. C. Sales, D. Mandrus, M. H. Pan, J. Zhang, R. Jin, and E. W. Plummer, Phys. Rev. Lett. **103**, 076104 (2009).

[9] F. Massee, Y. Huang, R. Huisman, S. de Jong, J. B. Goedkoop, and M. S. Golden, Phys. Rev. B **79**, 220517 (2009).

[10] Y. Yin, M. Zech, T. L. Williams, X. F. Wang, G. Wu, X. H. Chen, and J. E. Hoffman, Phys. Rev. Lett. **102**, 097002 (2009).

[11] T.-H. Kim, Z. Wang, J. F. Wendelken, H. H. Weitering, W. Li, and A. P. Li, Rev. Sci. Instrum. **78**, 123701 (2007).

[12] T.-H. Kim, J. F. Wendelken, A. P. Li, G. Du, and W. Li, Nanotechnology **19**, 6 (2008).

[13] L. Reimer, *Scanning Electron Microscopy: Physics of Image Formation and Microanalysis* (Springer-Verlag, Berlin, 1985).

[14] J. P. Eisenstein, G. W. Swift, and R. E. Packard, Phys. Rev. Lett. **43**, 1676 (1979).

[15] J. L. Cardoso and P. Pereyra, Phys. Rev. B **61**, 6360 (2000).

[16] M. Tinkam, *Introduction to Superconductivity* (McGraw-Hill, New York, 1996).

[17] R. T. Gordon, N. Ni, C. Martin, M. A. Tanatar, M. D. Vannette, H. Kim, G. D. Samolyuk, J. Schmalian, S. Nandi, A. Kreyssig, A. I. Goldman, J. Q. Yan, S. L. Bud'ko, P. C. Canfield, and R. Prozorov, Phys. Rev. Lett. **102**, 127004 (2009).

[18] D. P. Young, M. Moldovan, and P. W. Adams, Phys. Rev. B **70**, 064508 (2004).

[19] W. J. Skocpol, Phys. Rev. B **14**, 1045 (1976).

[20] S. Hasegawa, I. Shiraki, F. Tanabe, and R. Hobara, Current Applied Physics **2**, 465 (2002).

[21] K. M. Lang, V. Madhavan, J. E. Hoffman, E. W. Hudson, H. Eisaki, S. Uchida, and J. C. Davis, Nature **415**, 412 (2002).

[22] F. Ning, K. Ahilan, T. Imai, A. S. Sefat, R. Jin, M. A. McGuire, B. C. Sales, and D. Mandrus, J. Phys. Soc. Jpn. **78**, 013711 (2009).

[23] J.-H. Chu, J. G. Analytis, C. Kucharczyk, and I. R. Fisher, Phys. Rev. B **79**, 014506 (2009).

[24] W. D. Wise, K. Chatterjee, M. C. Boyer, T. Kondo, T. Takeuchi, H. Ikuta, Z. Xu, J. Wen, G. D. Gu, Y. Wang, and E. W. Hudson, Nat. Phys. **5**, 213 (2009).

Figure 1. (color online) Topographic images of cleaved surfaces of $BaFe_{1.8}Co_{0.2}As_2$ single crystal. (a) SEM image showing uniform contrast with some dark regions near the edge of crystal, as marked by a rectangular box. (b) Zoom-in SEM image showing domain structures. Marked regions by symbols and a dash line in (a) and (b) indicate where the transport measurements and composition probing are carried out. (c) STM images showing the cleaved *ab* plane of single crystal surface with an atomic step. Image size: $100 \times 100$ nm$^2$, set point: 500 mV, 200 pA, measured at room temperature. Inset: quasi-atomic resolution STM image of the cleaved surface.

Figure 2. (color online) Resistance (*R*) versus temperature (*T*) measured at four different sample locations as marked by symbols in Fig. 1. The inset shows the measurement configuration of four STM probes. The separation between two voltage probes is 3 μm.

Figure 3. (color online) (a) Temperature dependence of critical current at different sample locations. A 0.1 μV voltage cut-off, corresponding to an electric field criterion of $3 \times 10^{-4}$ V/cm, was used to define the $I_C$. The lines show fits to $I_C(T) = I_C(0)(1 - T/T_C)^n$. Inset: normalized $I_C$ (*T*) data superimposed onto the same curve of $I_C(T)/I_C(0) = (1 - T/T_C)^n$. (b) Critical current density derived from remanent magnetization of bulk crystal. The solid line fits to $j_c = j_c(0)(1 - T/T_c)^n$ of the corresponding data set.

Figure 4. (color online) Location dependent Co contents analyzed with WDS both in uniform and domain regions where local transport measurements were carried out.

Areal scan was averaged from the elemental maps along a 5 μm wide line. The line scan curve has been shifted downward by 0.1% for clarity.

Figure 5. (color online) Location dependent resistance measured with 4-probe STM at 21.1 K, which is just below bulk transition temperature (22 K). Non-zero resistances indicate that there is non-superconducting or normal metal phase at 21.1 K. Inset shows measurement configurations in domain regions.

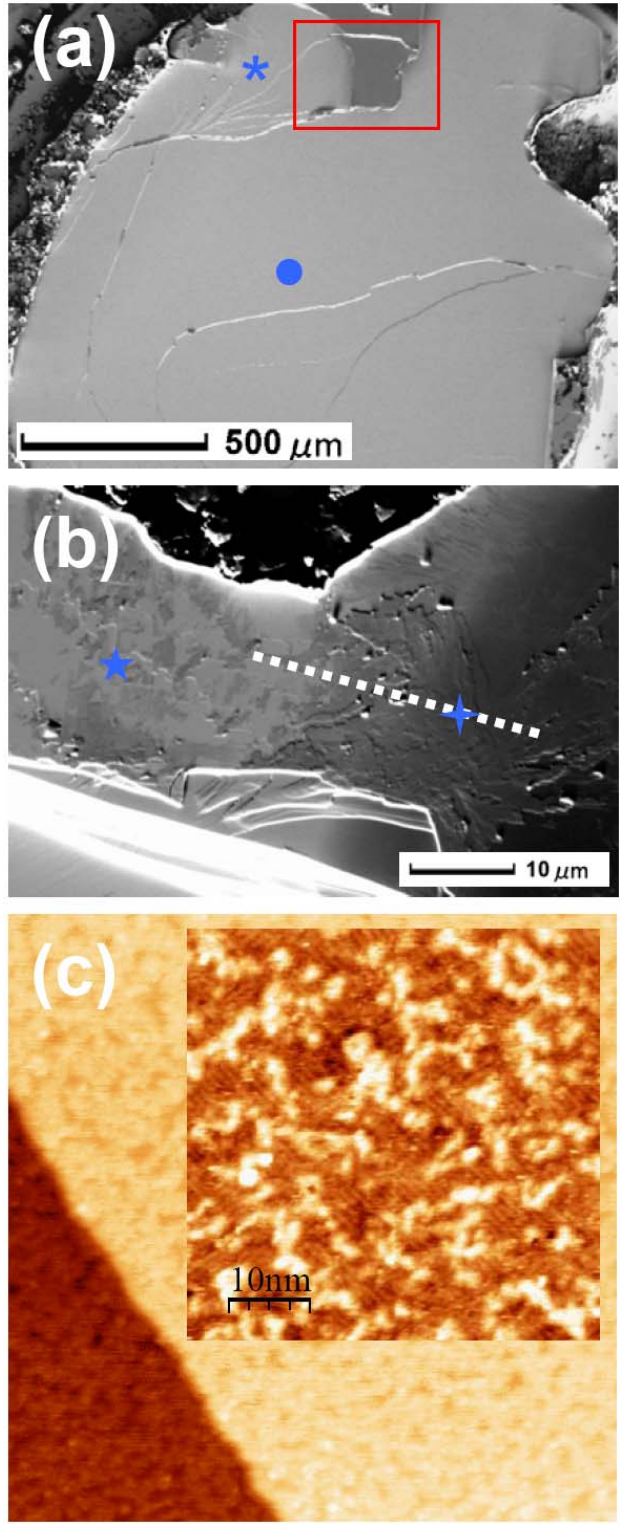

Fig. 1

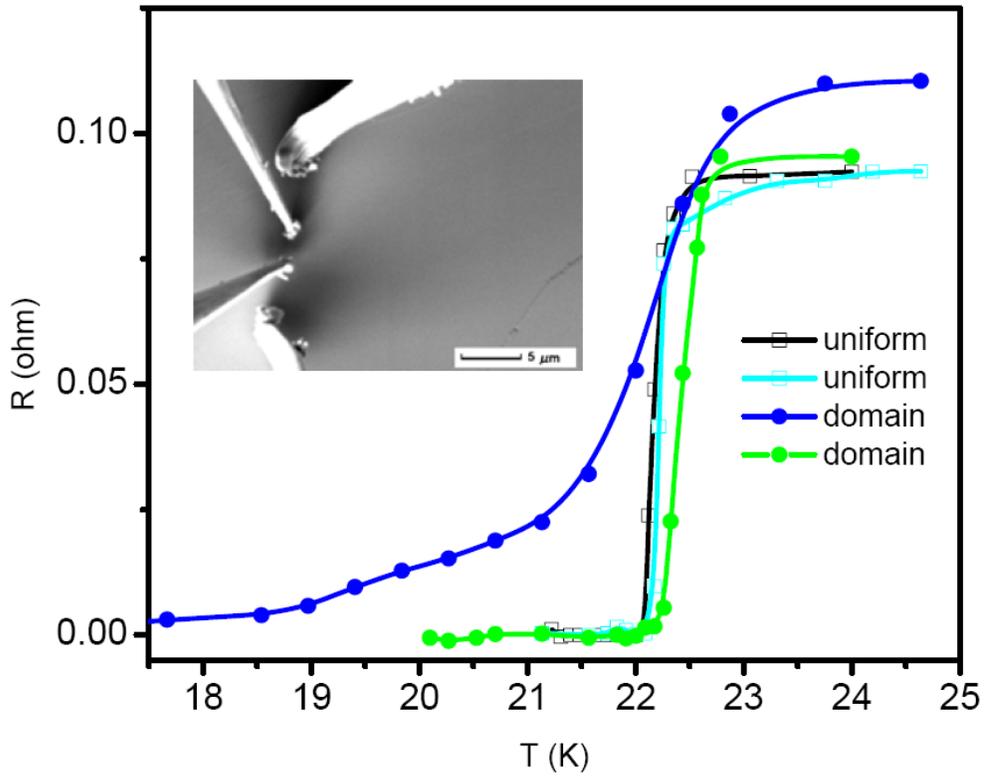

Fig. 2

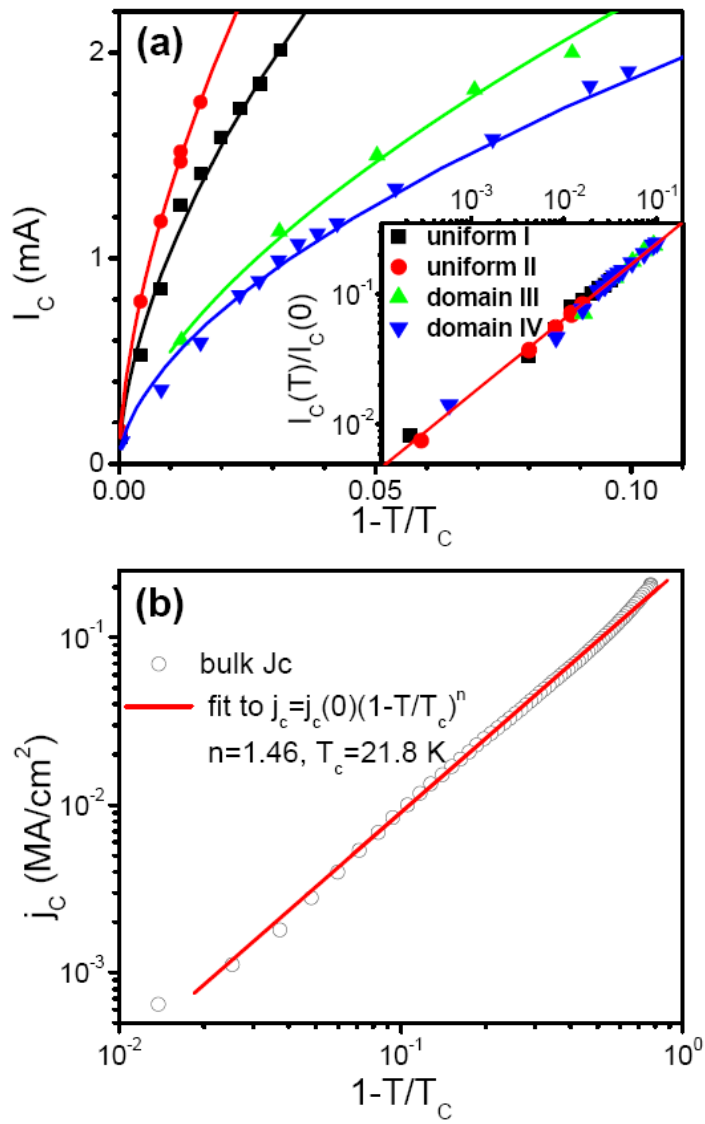

Fig. 3

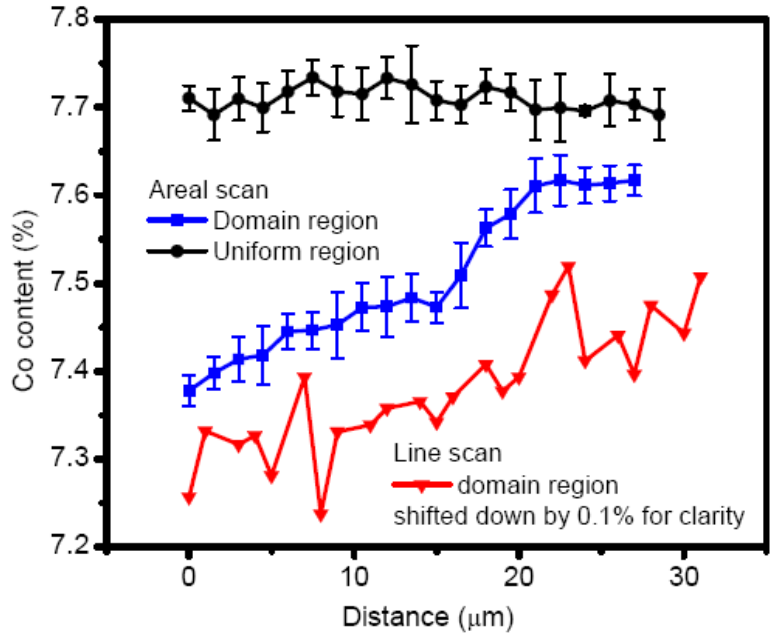

Fig. 4

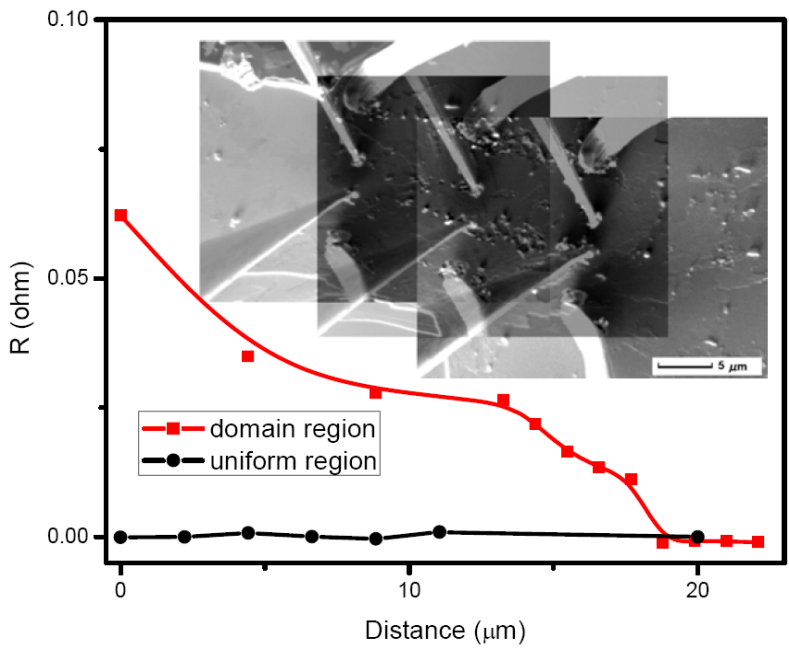

Fig. 5